\documentclass[a4paper,11pt]{article}
\pdfoutput=1 
\usepackage{jcappub}

\begin{document}
\title{Diffusion of massive particles around an Abelian-Higgs string}
\author{Abhisek Saha}
\author{Soma Sanyal}
\affiliation{School of Physics, University of Hyderabad, Gachibowli, Hyderabad, India 500046}

\emailAdd{saha\_abhisek@yahoo.com}
\emailAdd{sossp@uohyd.ernet.in}

\abstract {We study the diffusion of massive particles in the space time of an Abelian Higgs string. The particles in the early universe plasma execute Brownian motion. 
This motion of the particles is modeled as a two dimensional random walk
in the plane of the Abelian Higgs string. The particles move randomly in the space time of the string according to their geodesic
equations. We observe that for certain values of their energy and angular momentum, an overdensity of particles is observed close to the string. 
We find that the string parameters determine the distribution of the particles. We make an
estimate of the density fluctuation generated around the string as a function of the deficit angle. Though the thickness of the string is small, 
the length is large and the overdensity close to the string may have cosmological consequences in the early universe.}


\maketitle
\flushbottom

\section{Introduction}
Cosmic strings are linear topological defects generated in symmetry breaking phase transitions in the early universe due to the Kibble mechanism \cite{vilenkin}. 
Initially, they were very important as they were the only model apart from inflation that could give rise to seed density fluctuations \cite{cosmicstrings}. 
They subsequently lost their appeal when the data from the Cosmic 
Microwave Background Radiation (CMBR) experiments favored inflationary density fluctuations \cite{CMBR}. However, lately they have made a comeback. It was found 
that cosmic strings are generically formed at the end of the inflationary era. New studies have been done to detect these strings. Detailed simulations showing 
how these strings interact and grow have been performed. While initially the cosmic 
strings formed during the Grand Unified Theory (GUT) transitions were mostly studied, recently the Abelian Higgs string has also gained prominence. It is
often considered the prototypical field theory to study the constraints imposed on the string defects by the recent data from the Cosmic Microwave Background (CMB) 
\cite{CMB2}. Though density fluctuations from cosmic strings are a subdominant component of structure formation, they do give rise to prominent signatures in position
space maps. Recently, various signatures related to cosmic strings have been discussed in the literature, almost all of them relate to density fluctuations 
from the motion of these strings \cite{stringsigna1, stringsigna2}.

The Abelian Higgs cosmic string was first studied by Nielsen and Olesen \cite{neilsen}. Since then there have been detailed simulations of their growth and their 
structures \cite{abegrowth}. There have also been studies of particle motion in the metric of the cosmic strings. Generally if the strings are 
infinitely thin, they correspond to a flat conical metric. Geodesics on such metrics are straight lines. Recently, geodesics of particles around an
Abelian Higgs strings with a finite core has also been studied \cite{hartman}. The authors find that bound orbits exist for massive particles for certain string 
parameters.

As cosmic strings move through the early universe plasma, wakes are generated behind them due to the presence of the deficit angle \cite{wakes}.
Wakes have mostly been studied for an infinitely thin cosmic string, however other simulations have shown that the wakes due to wiggly cosmic strings can 
give rise to shocks and generate magnetic fields \cite{vachaspati}. Wakes are overdensities generated by the motion of the strings. In this work, we 
suggest that overdensities can also be generated by massive particles becoming trapped close to an Abelian Higgs cosmic string. A single particle moves 
along the geodesic given by the underlying metric. However, if there are a large number of particles (as it would be in the case of the early universe plasma),
the particles would collide with each other and execute a Brownian motion. A Brownian motion on a flat metric will yield no overdensity. But if we replace the 
flat metric with the metric of the Abelian Higgs model and model the Brownian motion of the massive particles around the Abelian Higgs string; then
we find an overdensity of particles close to the string.

In this work we are considering massive particles.   In the cosmological scenario, this means baryons, WIMPS and any other massive exotic dark matter particle. In these simulations, we do not specify any particular particle but study the clustering of a collection of massive particles. 
As we do not consider any interactions between the particles, we model them moving around a static Abelian Higgs string as a random walk problem. 
We briefly explain why we are using a random walk to model the motion of the particles. Generally, the early universe plasma is a neutral plasma consisting 
of many particles. Even though these are charged particles, there is negligible electric field in the background plasma. Particles moving in this plasma, 
randomly collide, with other particles which are also moving in the plasma. The effect of many successive elastic collisions leads to a Brownian motion. 
This Brownian motion is modeled as a random walk of $N$ particles around a cosmic string where $N$ is taken to be large. As mentioned before in between collisions, 
the particles move according to their geodesic equations. In flat space time, this means that they move in straight lines. In the case of the Abelian 
Higgs string the geodesic equations are obtained from the cosmic string metric. The particles only undergo elastic collisions. We change the direction of 
their velocities but the magnitude of the velocity is kept constant. We give the details of the model in the next section. We find that the particles do start 
clustering around the string. This means that even for a static string we get some density fluctuations. We find the order of magnitude of the fluctuations 
as a function of the deficit angle of the string. We find that  the clustering depends crucially on the deficit angle. As the deficit angle increases, the 
particles start clustering closer and closer to the string. However, when it becomes close to $2 \pi$, the particles cease to come any closer. 
Since we are interested in the clustering phenomenon, we study the strings with deficit angle less than $2 \pi$. We have included a brief discussion on the 
simulation results for angles greater than $2 \pi$.  


In section II we present our model in detail. In section III, we discuss our choice of simulation parameters for the random walk problem. In section IV, 
we present the results and discuss the cosmological consequences of the density fluctuation. Finally we present our conclusions in section V. 

\section{The Model}

The Abelian Higgs model has been used as prototypical model to model cosmic strings in the early universe. The Lagrange density for the model can be 
written as 
\begin{equation}
 \mathcal{L} = - \frac{1}{4} F_{\mu \nu} F^{\mu \nu} + D_{\mu} \phi D^{\mu} \phi^{*} - \frac{\lambda}{4} (\phi \phi^* - \eta^2)^2 
\end{equation}
Here $D_{\mu} \phi$ is the covariant derivative given by $D_{\mu} \phi  = \nabla_\mu \phi - i e A_\mu$ and 
$F_{\mu \nu} = \partial_\mu A_{\nu} -\partial_\nu A_{\mu}$ is the Abelian field strength. $\phi$ is a complex scalar field with vacuum expectation value $v$. 
Now this Lagrange density allows stable vortex solutions which in 3 - dimensions
leads to the Abelian Higgs strings. The gravitational effect of this string is modeled by coupling the Abelian Higgs model
minimally to gravity. The action is then given by \cite{hartman}, 
\begin{equation}
 S=\int d^{4}x \sqrt{-g}(\frac{1}{16\pi G}R + \mathcal{L} ), 
\end{equation}
where $R $ is the Ricci scalar. 
The cosmic string is cylindrical in shape and the most general line element obeying all the symmetry properties is given by 
\begin{equation}
 ds^2 = N^2(\rho) dt^2 - d\rho^2 - L^2(\rho) d\phi^2 - N^2(\rho) dz^2  
\end{equation}
The factors $L(\rho)$ and $N(\rho)$ are determined by the boundary conditions.They are related to the values of the fields at a distance $\rho$ 
from the axis of the string (the $z$ axis in this case).
The equations of motion for the two fields had been solved for large $\rho$ by Neilson and Olesen. An exact solution was obtained by de Vega et. al. 
Using the Neilson and Olesen ansatz, 
\begin{eqnarray}
 \phi(\rho, \phi) \propto  f(\rho) e^{-i n\phi}   \\
   A(\rho, \phi)  \propto - \frac{A(\rho)}{\rho} 
\end{eqnarray}
  There is a magnetic field along the z-axis and its value too depends on the scalar and the vector potentials.
The metric corresponds to a cylindrical metric with a deficit angle. The deficit angle far from the core of the 
string is proportional to the energy per unit length of the string. If the vacuum expectation value {\it(vev)} of the Higgs field is sufficiently large, then 
the deficit angle can be larger than $2 \pi$. These are called super-massive string. 

Further, the Lagrangian of the model is usually rescaled and written in terms of two dimensionless 
constants. 
\begin{equation}
 \gamma = 8\pi G \eta^{2}\ ,\   \beta = \frac{\lambda}{e^{2}}
\end{equation}
The deficit angle depends upon these two constants. The cosmic string has a finite width with a core of magnetic flux as well as a scalar core.
The width of these cores are the inverse of the gauge boson mass and the Higgs mass respectively. The Bogomolny limit occurs when the Higgs mass is equal 
to the gauge boson mass. This happens for $\beta = 2$. For $\beta < 2$, $\Delta < 4 \pi \gamma $ and for $\beta > 2$,  $\Delta > 4 \pi \gamma $. The latter
corresponds to the super massive strings. For $\beta = 2$, the deficit angle is given by, 
\begin{equation}
 \Delta = 4 \pi \gamma  
\end{equation}

Now the particles in the plasma move around and collide randomly against one another.  
We assume that there are no other forces on the particles in the plasma except the gravitational effect of the Abelian Higgs cosmic string. We do not 
consider any gravitational attraction between the test particles.
The particles move according to their geodesic equations and collide with one another randomly. The collisions are considered to be elastic collisions. 
The motion of the particles are therefore similar to the Brownian motion of particles in the metric of an Abelian Higgs string. So there is no 
change in the magnitude of the velocity, only the direction of the particles can change. Hence,we model the diffusion of particles around a cosmic string as 
a random walk problem. We do not allow particles to overlap and we also confine ourselves to a two-dimensional random walk in the $\rho-\phi$ plane of the cosmic 
string.

We can obtain the geodesic equations in the $\rho-\phi$ plane of the cosmic string from the general equation, 
\begin{equation}
\frac {d^2 x^{\mu}}{d \tau^2} + \Gamma_{\rho \sigma}^{\mu} \frac{dx^{\rho}}{d \tau} \frac{dx^{\sigma}}{d \tau} = 0 
\end{equation}
where $\Gamma_{\rho \sigma}^{\mu} $ is the Christoffel symbol and $\tau$ is the affine parameter. The time like geodesics in this case correspond to the 
proper time. The geodesic Lagrangian for a massive particle will then be, 
\begin{equation}
 N^2 \left(\frac{dt}{d \tau} \right)^2 - \left(\frac{d\rho}{d \tau} \right)^2 - L^2 \left(\frac{d\phi}{d\tau} \right)^2 - N^2 \left(\frac{dz}{d\tau} \right)^2 = 1 
\end{equation}
The constants of motion are the total energy $E$ given by $E = N^2 \frac{dt}{d \tau}$ ; the component of the angular momentum along the $z$ axis $L_z (= L^2 \frac{d\phi}{d\tau}$),  
and the linear momentum in the $z$ direction ($p_z = N^2 \frac{dz}{d \tau}$). Using these constants of motion we can get the geodesic equation in the 
$\rho - \phi$ plane as,  
\begin{equation}
 \frac{d^{2}\rho}{d\tau^{2}}= (p_{z}^{2}-E^{2})\frac{N^{'}}{N^{3}}+L_{z}^{2}\frac{L^{'}}{L^3} ,  
\end{equation}
and 
\begin{equation}
 \frac{d^{2}\phi}{d \tau ^{2}}= -2L_{z} \frac{L^{'}}{L^{3}}  \frac{d\rho}{d\tau} . 
\end{equation}

As mentioned before the space time of an Abelian Higgs string has a deficit angle. In this paper we have only considered cosmic strings with deficit angle $\Delta < 2 \pi$. We do not consider the supermassive strings. 
The deficit angle is implemented in the boundary 
conditions of the random walk problem. Unlike the flat space, where the value of $\phi$ varies between $0$ to $2 \pi$, in our simulation, the value of $\phi$ varies 
between $0$ to $(2 \pi - \Delta)$.  Generally, the set of geodesic equations
for the Abelian Higgs string can be solved numerically for the following boundary conditions, 
\begin{equation}
 f(0)= 0, \quad N(0)= 1,  
  \quad N^{'}(0)= 0, \quad L(0)= 0,\quad L^{'}(0)= 1
\end{equation}
These come from the requirement of regularity of the origin \cite{brihaye}. 
Since particles are neither created nor destroyed during the time period of the simulation, we check that the total number of particles are conserved at each time step.
Between two collisions, the particles move according to their geodesic equations. The equations are solved numerically using a standard Runge Kutta routine. We initially 
check the code for a standard random walk in flat space, then we modify the equations of motion and run the code for $N$  particles where $N$ is a large number. 
In the next section, we discuss the simulation parameters in detail.

\section{Simulation Parameters}

We use a random number generator to assign the position of $N$ particles in the $\rho - \phi$ plane of the string. The cosmic string is positioned right at the center. 
We have fixed the horizon size to be much greater than the random walk step size. If $d_H$ is the size of the horizon at time $t$, then the units in our simulation 
correspond to $x$, where $x = d_H/100 $. The random walk step size is given in these units. Typical random walk steps are less than $0.1 x$. We have also kept the 
velocity of the particle flexible. The magnitude of this velocity is given initially and remains fixed through out the simulation. However, the direction is 
changed randomly after every random walk step or when it collides with another particle. This is done by checking that there are no overlapping particles. 
We have run the simulation for a large number of particles starting from $N = 1000$. Our final graphs and results are from $N = 10,000$ particles, which are 
initially distributed between various valuse of $\rho$ and $\phi$ varying between $0$ to ($2 \pi - \Delta$). $\Delta$ being the deficit 
angle which we keep as a variable parameter in our simulations. Between collisions or between the change of direction, the particles move according to the geodesic 
equations obtained in the previous section.  We take snapshots after every ten time steps and store the data. The data is then binned in the x-y coordinates.

We are interested to see if particles diffusing around an Abelian Higgs string have any tendency of clustering around the string due to the presence of the magnetic 
and scalar field cores.  The strings have a small width but can have a considerable length. The nature of the particle motion is 
determined by an effective potential in which the particles move. The effective potential is given by \cite{hartman}, 
\begin{equation}
 V_{eff} (\rho) = \frac{1}{2} \left [ E^2 \left(1 - \frac{1}{N^2} \right ) + \frac{p_z^2}{N^2} + \frac{L_z^2}{L^2}\right] 
 \label{veff}
\end{equation}
Our values of $\beta$ and $\gamma$ are chosen such that the effective potential has a well defined minima. The result is that $\beta$ is always less than two. 

\section{Results and Discussion}
In this section, we present the results of our simulations. Fig. 1 gives the initial particle distribution for $N = 10,000$ particles. After that we have 
taken snapshots after every $10$ timesteps. We find that the particles that were randomly spread around the cosmic string start to move closer to the magnetic 
core of the string as time goes on. We have done the simulations for different values of the deficit angle. We present here some selected snapshots of the 
particles at certain intervals to show how the clustering occurs.  For figure 1, we show the intial distribution of $N = 10000$ particles around a cosmic string with parameters $E = 1.083$, $L_z^2 = 0.025$, 
$p_z= 0.02$ and $\gamma = 0.32$. 
\begin{figure}
\includegraphics[width = 86mm]{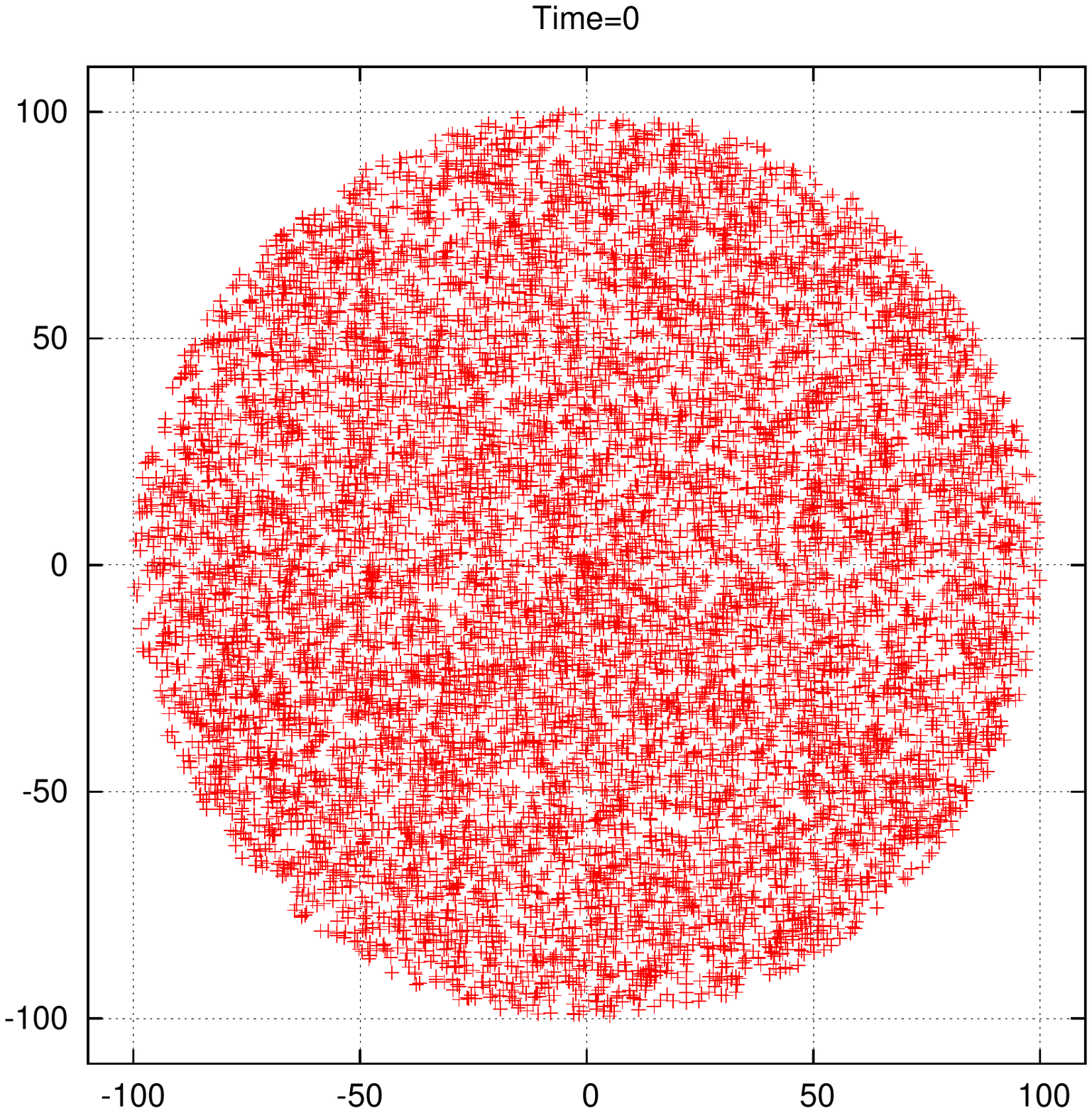}
\caption{Initial distribution of particles around the cosmic string. }
\end{figure}
Fig. 1 gives the full range plot. However, in the full range plot, it is difficult to discern the clustering effect visually. We therefore focus on a shorter
range closer to the cosmic string. The position of the string is right at the middle of the graph. 
There are fewer particles in the graph and it is easier to visualize  the diffusion of the particles towards the string in these plots.  
Fig. 2 now gives the initial position of the particles in the range $[-10:10]$; fig. 3 is taken after 10 steps and fig. 4 is after 
500 time steps. We see that the number of particles in this range is increasing. However, around $500$ timesteps the simulation reaches an equilibrium 
distribution. Even though the position of the particles change, the density within a certain range remains the same.  

\newpage

\begin{figure}
\includegraphics[width = 86mm]{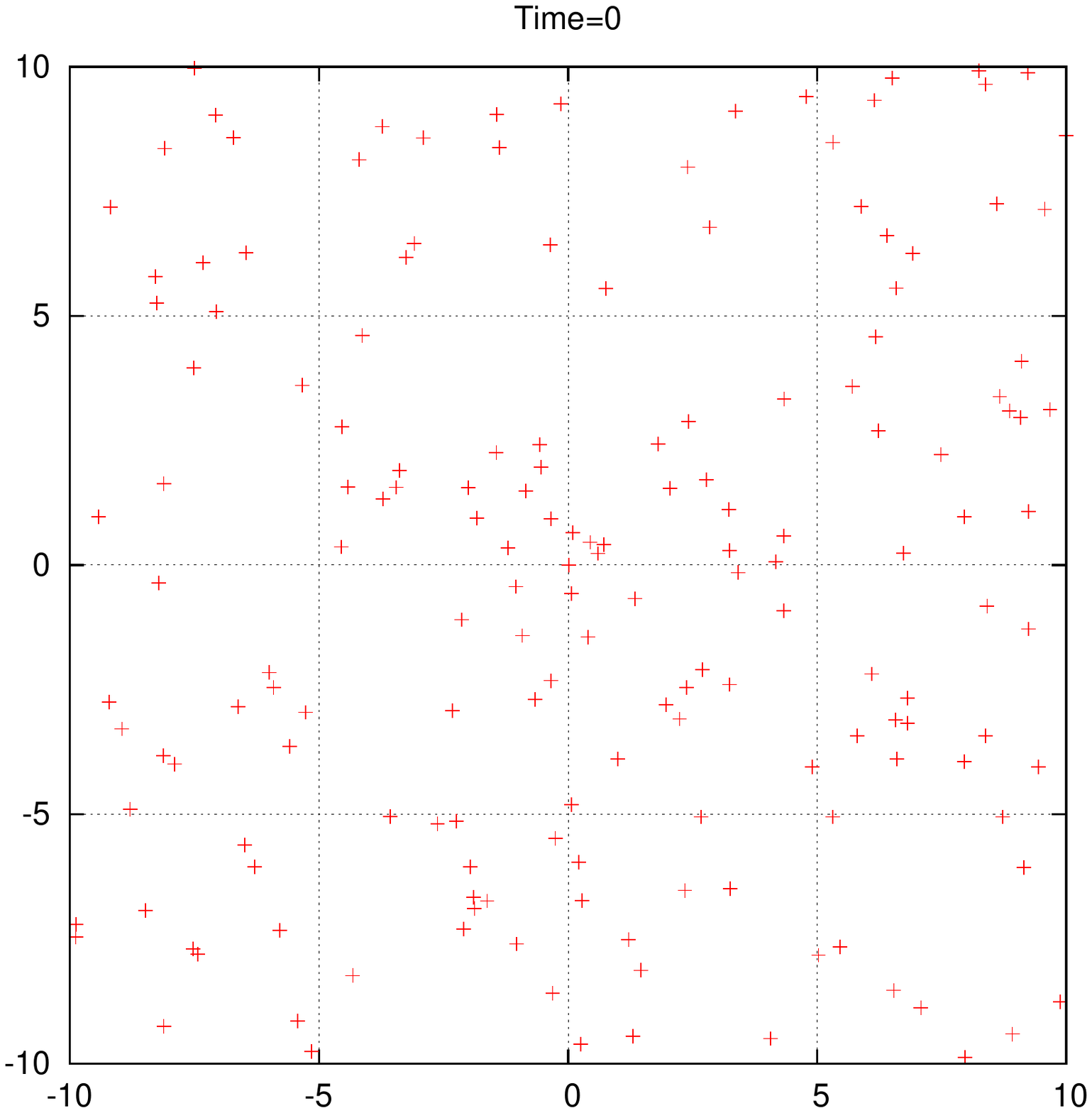}
\caption{Particles in the range [-10:10] around the cosmic string. }
\end{figure}
\begin{figure}
\includegraphics[width = 86mm]{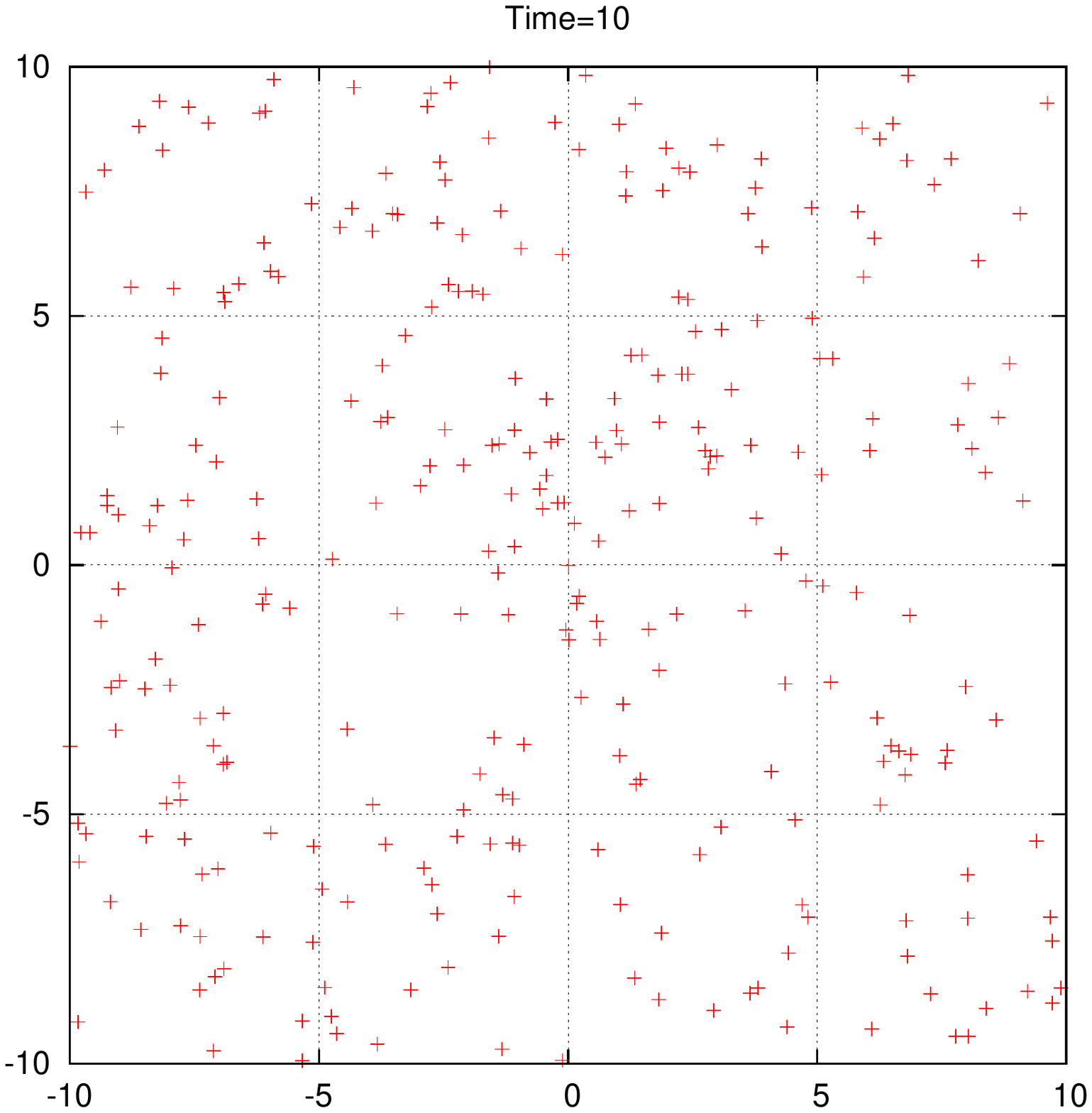}
\caption{The distribution of particles after 10 timesteps. }
\end{figure}
\begin{figure}
\includegraphics[width = 86mm]{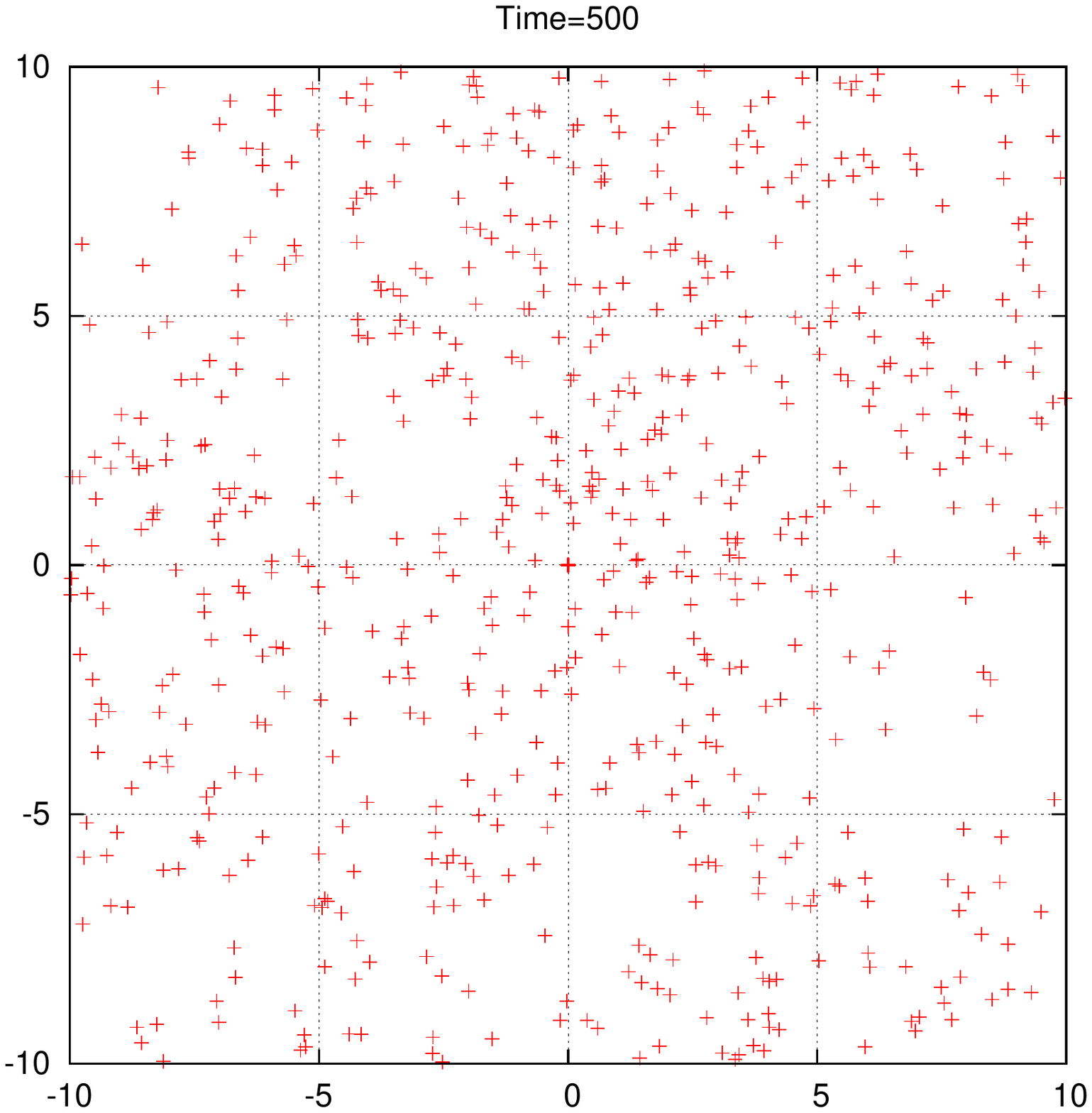}
\caption{The distribution of particles after 500 timesteps. }
\end{figure}

\newpage

Though it appears that particles are moving towards the core of the string, it is difficult to make any quantitative estimates from the scatter plots. 
Since we would like to make a quantitative estimate of the overdensity generated near the string, we have binned the data in the X - Y plane.
Intially we binned the data for a random walk in flat space time with $N = 10,000$ particles scattered in the range $[-100:100]$. The binned data 
showed a Gaussian with a broad peak. This is what we took as the background density $\rho_0$. We then binned the data for the case of the Abelian - Higgs 
string for different values of the deficit angle. The binned data clearly shows an increase in the density near the center of the string.

In fig.5 we show the density distribution around a cosmic string with parameters $E = 1.083$, $L_z^2 = 0.025$, $p_z= 0.02$ and $\gamma = 0.15$. For comparison, 
we have also plotted the Gaussian we obtained for the case when the cosmic string was not there. 

\begin{figure}[b]
\includegraphics[width = 86mm]{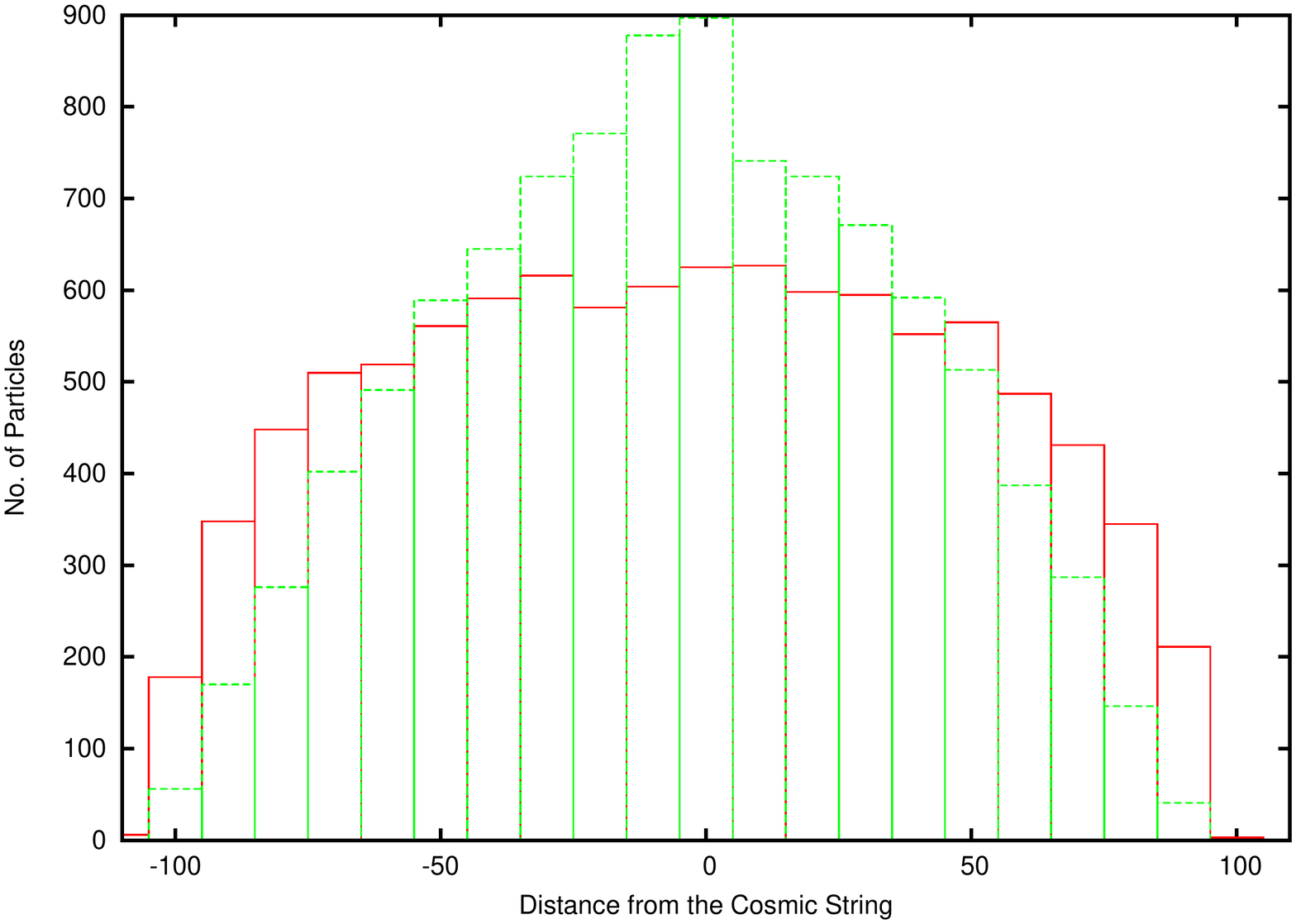}
\caption{Histogram of a density distribution around the cosmic string. The solid (red) lines denote the distribution in the absence of the cosmic string, 
while the dashed (green) line denotes the density distribution for an Abelian Higgs string with the parameters $\gamma = 0.15$, $E = 1.083$, $L_z^2 = 0.025$, $p_z= 0.02$ }
\end{figure}
We have varied the parameters of energy $E$, angular momentum $L_z$, as well as $p_z$, however we find that the clustering is strongly sensitive to the parameter 
$\gamma$. This is expected, as it is this parameter which determines the width of the magnetic core of the string. Since it is the clustering effect
we are interested in hence we keep the other parameters constant and vary only $\gamma$. So the figures presented are for $E = 1.083$, $L_z^2 = 0.025$ and $p_z= 0.02$.
In fig. 6 and fig. 7 we give the density distribution for two different $\gamma$ values.
\begin{figure}
\begin{minipage}{0.45\textwidth}
\centering
\includegraphics[width = 50mm]{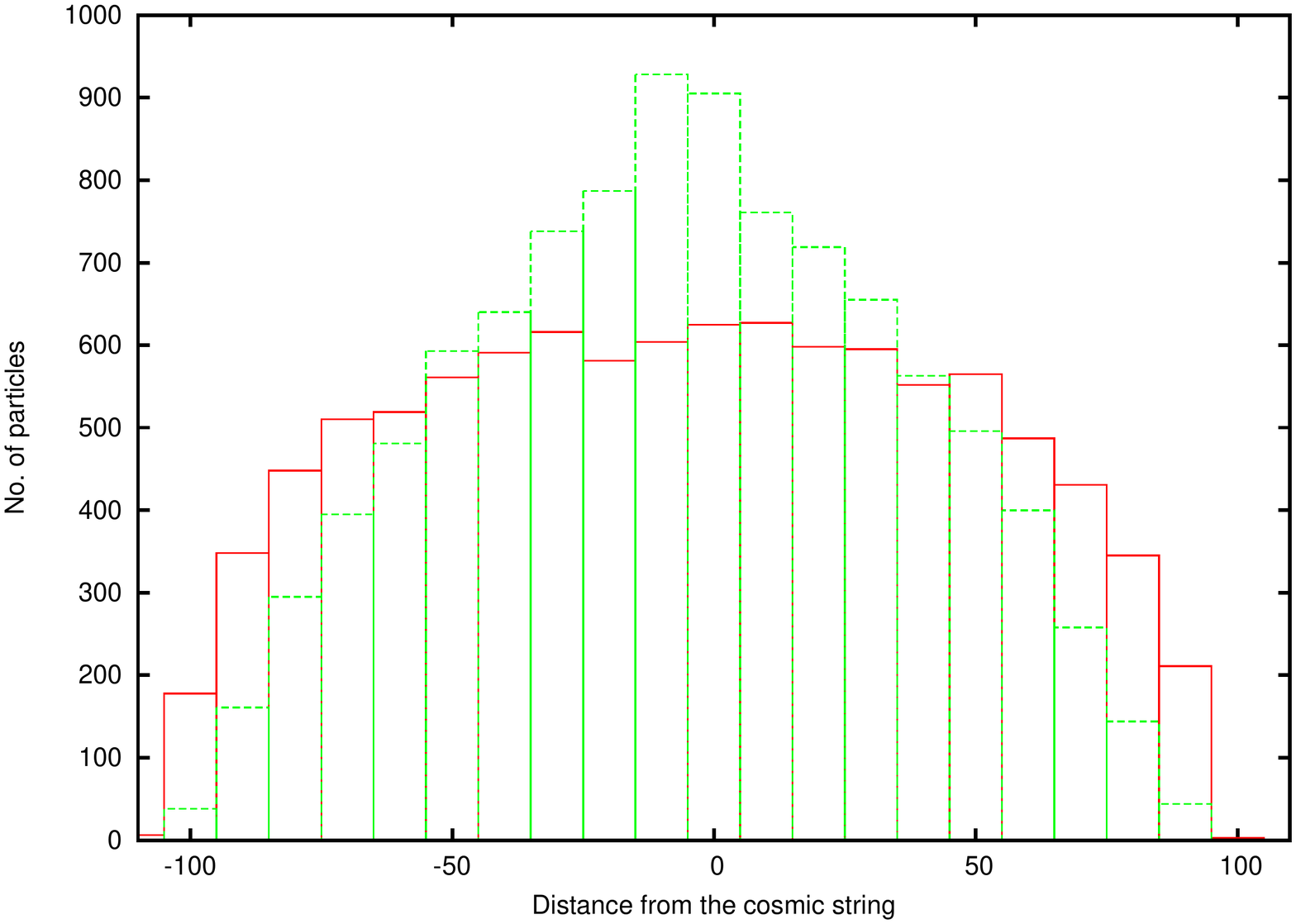}
\caption{Histogram for $\gamma = 0.24$}
\end{minipage}%
\begin{minipage}{0.45\textwidth}
\centering
\includegraphics[width = 50mm]{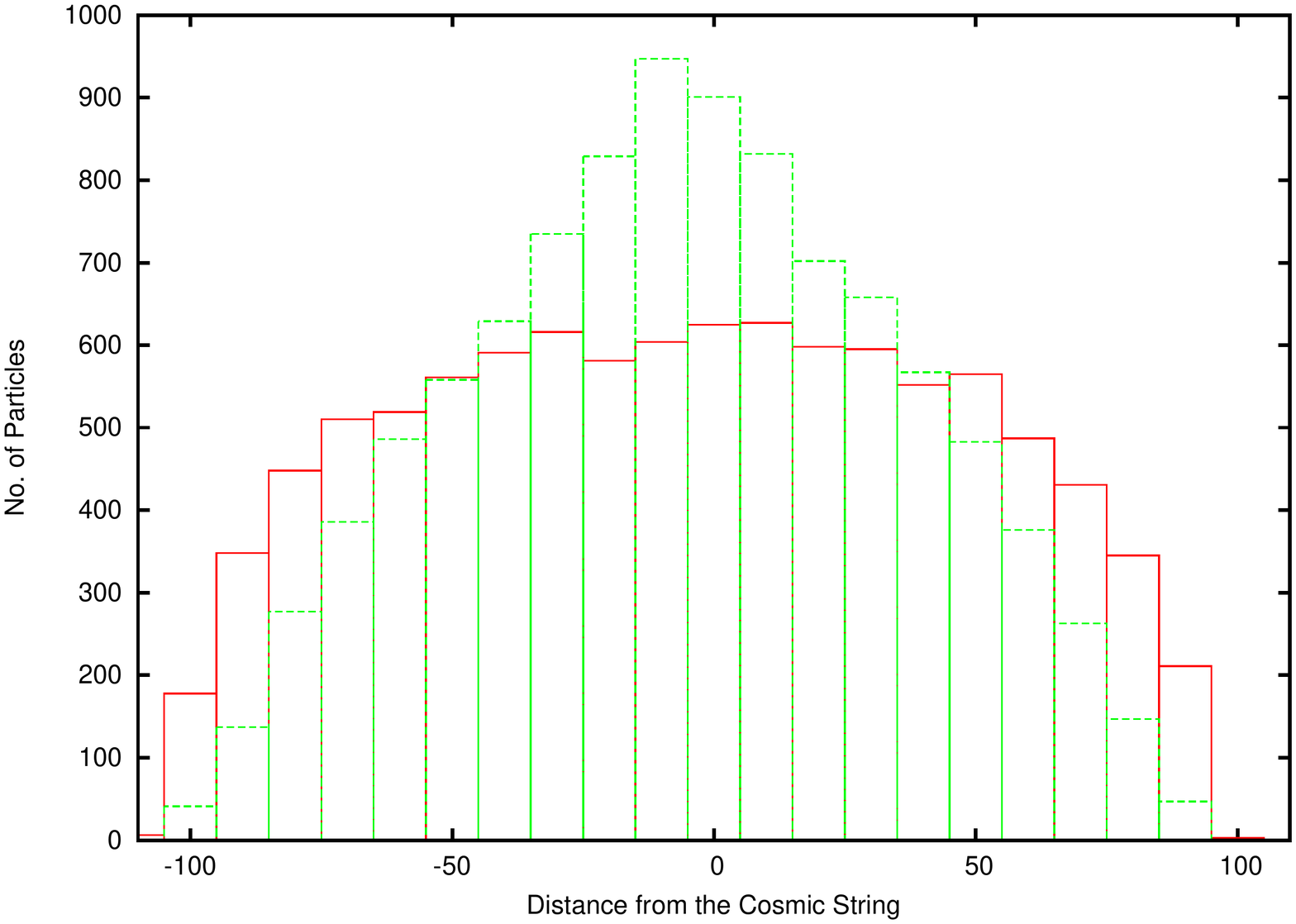}
\caption{Histogram for $\gamma = 0.40$}
\end{minipage}
\end{figure}

As mentioned before, we have denoted the average distribution of the particles in the absence of the string as $\rho_0$. We notice from the binned data that  
the particles distributed about the cosmic string metric with deficit angle $\Delta$ will give different average distribution $\rho_1, \rho_2,.....$ etc 
for different values of $\Delta$. All the binned distribution are Gaussian, only the peak becomes sharper as the deficit angle is increased. So the average
values are obtained by determining the half maxima of the distribution. Now we define $\delta \rho$ as the difference between $\rho_0$ and the average 
distributions $\rho_i$ where each $i$ corresponds to a different deficit angle. So for any deficit angle, we can calculate, $\frac{\delta \rho}{\rho_0}$. 
Finally we plot the change in the density distribution $\frac{\delta \rho}{\rho_0}$  against $\gamma$ in fig 8. We see that the final density contrast is 
quite high, nearly close to $0.5$. However, the distribution always remains a Gaussian. Though the density contrast tends to become quite high, we have not 
considered any non-linear effects. Usually, non-linear effects cause departures from Gaussianity, since 
the distribution always remains Gaussian, hence we do not consider any non-linear effects at this point. One reason for this may be, 
we are considering only static strings here. In the future, we plan to look at the accretion of particles around a moving string, and include the non-linear evolution of the density 
fluctuations. 
\begin{figure}[b]
\includegraphics[width = 86mm]{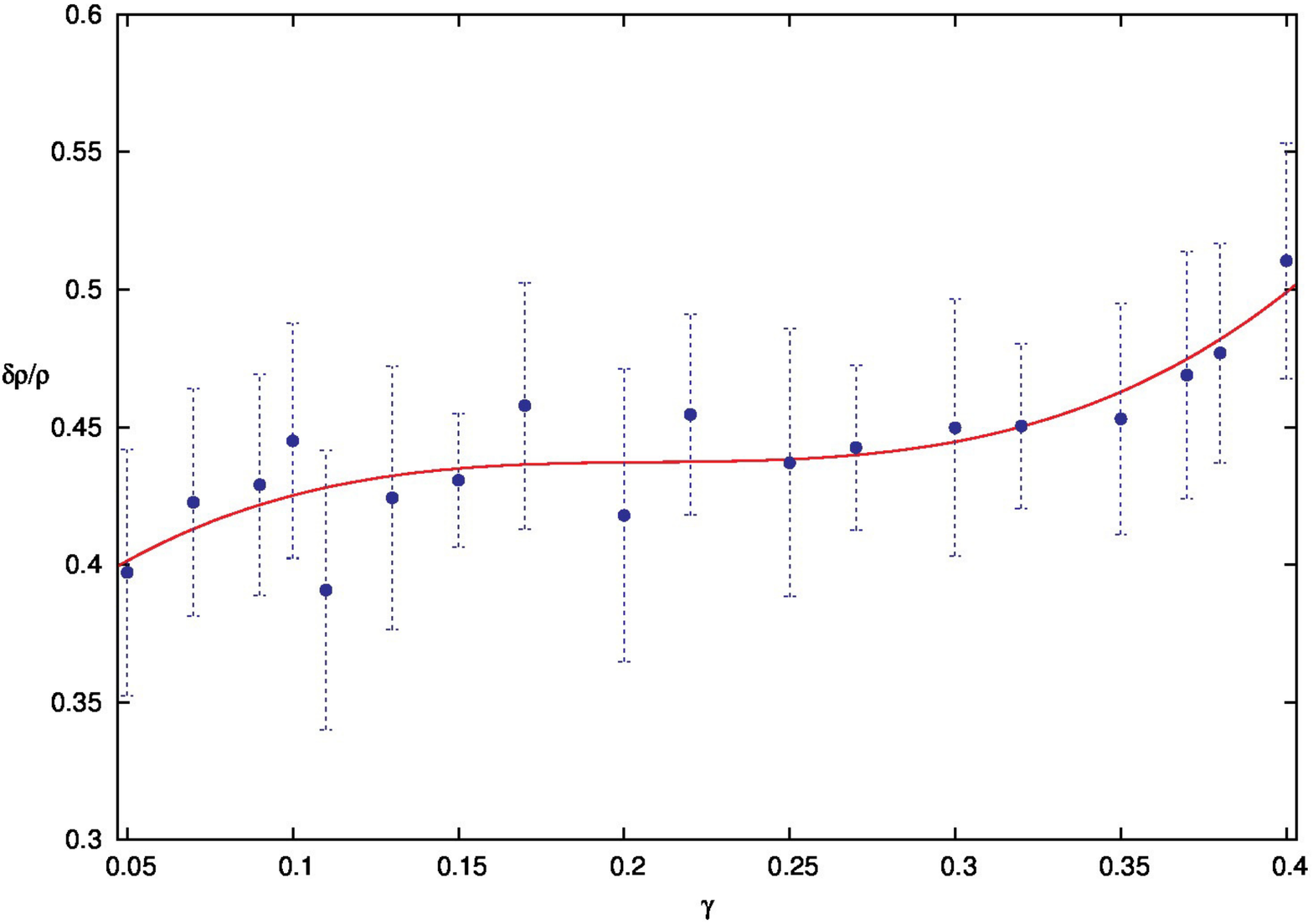}
\caption{Variation of over density with $\gamma$. The deficit angle is directly proportional to $\gamma$ }
\end{figure}

In fig. 8, we have kept the $\gamma$ values between $0.1 - 0.4$ since a significant change occurs for $\gamma \approx 0.45$. We find that the 
particles are repelled from the center and become confined between an annular ring around the string. The histograms therefore show a dip in the center with two symmetrical peaks on either side.
For such $\gamma$ values, the deficit angle is very close to $2 \pi$.
The nature of the density distribution clearly depends on the deficit angle. The deficit angle is directly related to $\gamma$. As $\gamma$ increases,
the deficit angle increases too. The deficit angle is related to the width of the magnetic flux core and the scalar core. From fig.8, we see that initially, 
as long as the deficit angle remains between $0 - \pi$, the increase in the overdensity is gradual. But, as the deficit angle increases beyond $\pi$, there 
is a sharper increase in the clustering of the particles. Beyond $2 \pi$, no clustering of particles is observed. 

Initially, we put boundary conditions such that the total number of particles 
are conserved. We than removed the boundary condition and allowed the particles to move out of the space, if necessary. For $\gamma < 0.4$, our results 
remained the same but we found that for $\gamma > 0.5$, there 
are no particles left in the vicinity of the string cores after some time steps. This occurs because for  $\gamma > 0.5$, the deficit angle becomes greater 
than $2 \pi$, hence no clustering is observed. In fact, the particles in this realm quickly move out of the space close to the string cores.

We have also done the statistical error estimation for the different deficit angles. We find that the average value of $\frac{\delta \rho}{\rho_0}$ generally
converges in about  $20$ to $25$ runs for a particular deficit angle. We have obtained the standard deviation, and plotted the error bars for each deficit 
angle. The error bars would reduce if the number of particles are increased. Finally, we have done a best fit with the data points obtained. Though some of 
the points diverge from the solid line (best fit line), the line is within their error bars. The maximum size of the error bars are about $0.09$ as shown in 
Fig. 8.

Our results seem to be consistent with previous work on Abelian Higgs strings. In  ref.\cite{hartman}, where bound geodesics were observed, they found that the 
nature of the bound orbits depended on the nature of the potential given in equation \ref{veff}. Since both particle trapping and bound orbits depend upon the
presence of a minima in the potential, it seems that the clustering of particles would occur for potentials where bound geodesics have been observed.  It has 
been observed that the maximum radius of the bound orbit around an Abelian Higgs string decreases with increasing $\gamma$. Fig. 8 in 
ref.\cite{hartman} shows 
that the maximum radius for $\gamma = 0.36$ is approximately $2$, for $\gamma = 0.42$ is approximately $1.5$ and for $\gamma = 0.48$ is less 
than 1.5. Beyond $\gamma > 2 \pi$, there are no bound orbits. Both these are reflected in our results, we find that the overdensity close to the string increases with 
an increase in $\gamma$ and there is no clustering effect observed when $\gamma$ becomes greater than $ 2 \pi$.

Though Abelian Higgs strings are the simplest model for gauged strings we find that the presence of magnetic and the scalar cores lead to clustering of 
massive particles around the string. A more cosmologically interesting but challenging model would be the electroweak model which gives us electroweak strings\cite{electroweak}.
Our simulations also seem to  indicate that the presence of bound orbits around a cosmic string are a strong signal that density fluctuations may be generated 
in the vicinity of the string. On a larger scale, it may be applicable to other systems where bound geodesics have been observed for massive particles. 
It would also be interesting to study the collective motion of particles for supermassive strings.

\section{Conclusions}
We have modelled  particle diffusion in the vicinity of a static cosmic string as a two dimensional random walk problem in the space-time of an Abelian Higgs 
cosmic string. We find that the particles start clustering around the cosmic string. This means that we get density fluctuations around a stationary cosmic 
string. We find that the density fluctuations obtained are not too sensitive to the energy or the angular momentum of the particles. The density fluctuations 
depend on the deficit angle of the space around the cosmic string. Though they are small for small deficit angles, they increase and   
become of the order of $0.4-0.5$ for large deficit angles. As the deficit angle increases, the density 
of particles close to the string increases. This continues till the we reach angles of the order of $1.8 \pi$. Beyond this, the particles start moving away 
from the core of the string. As the deficit angle goes beyond $2 \pi$, particles start to diffuse away from the vicinity of the string. So there is no clustering
effect observed beyond  $\Delta = 2 \pi$.

Our results are consistent with the conclusions reached previously in ref.\cite{hartman}. They have studied the geodesics of massive particles around Abelian - Higgs 
cosmic strings. They had found that the maximal radius of the bound orbits decreases with an increase in the deficit angle. In our case, this translates to the particles 
being trapped closer and closer to the string. In our case, we do not look at the super-massive strings. Hence, we never go beyond the Bogomolny limit. Since the 
clustering seems to be correlated to the presence of the bound orbits, it is quite possible that the collective motion of massive particles
will give rise to density fluctuations for bound orbits observed in other systems also \cite{ulbricht}.

As is well known cosmic strings are not static. Connected strings grow as a network and individual strings move through the plasma.  As the strings move,  
wakes are formed behind them. These wakes have several consequences in the early universe \cite{layek,duple,cunha}. The structure of these wakes have been 
studied for infinitely thin  cosmic strings and wiggly cosmic strings. The overdensity behind the string is related to the deficit angle subtended by the 
string. In the case of the Abelian Higgs strings we believe that the wake structure will be modified by the clustering of particles around the string. It is 
quite possible that more particles will get trapped in the wake of the string and the overdensity will be enhanced. This may have several consequences for 
phase transitions in the early universe. We plan to address these and other issues in a future work.

\begin{center}
 Acknowledgments
\end{center} 
 A.S would like to acknowledge suggestions and advice from B. Bambah, R. Mohanta and E. Harikumar. 
 S.S would like to acknowledge discussions with Ajit M. Srivastava which led to the formulation of the current research problem. 
 The authors would like to thank Sanatan Digal for critical reading of the manuscript and his comments and advice which helped in improving the 
 article significantly.


\begin{thebibliography}{100}
 
\bibitem {vilenkin} Vilenkin, Alexander, and E. Paul S. Shellard. Cosmic strings and other topological defects. Cambridge University Press, 2000.


\bibitem{cosmicstrings}Hindmarsh,M., Stuckey, S. and Bevis, N. Phys Rev {\bf D.79}.123504, 2009 ;
Vachaspati, Tanmay, and Alexander Vilenkin. 
Physical Review Letters 67.9 (1991): 1057 ;
Brandenberger, Robert H., et al. 
International Journal of Modern Physics A 22.21 (2007): 3621-3642 ;
Avelino, P. P., et al. 
Physical Review Letters 81.10 (1998): 2008.

\bibitem {CMBR}K. Freese, and William H. Kinney.
 Journal of Cosmology and Astroparticle Physics 2015.03 (2015) 044;
 Bartolo, Nicola, S. Matarrese, and A. Riotto. 
 Advances in Astronomy 2010 (2010).
 
\bibitem{CMB2} J. Lizarraga, et al. 
Journal of Cosmology and Astroparticle Physics 2016.10 (2016): 042.
C. Dvorkin,  M. Wyman, and W. Hu. 
Physical Review D 84.12 (2011): 123519.

\bibitem{stringsigna1} N. Kaiser and A, Stebbins, 
Nature (London) 310, 391 (1984) ;
R. H. Brandenberger, R. J. Danos, O. F. Hernández, and
G. P. Holder, 
J.Cosmol. Astropart. Phys. 12 (2010) 028.

\bibitem{stringsigna2} O. F. Hernández, Y. Wang, R. Brandenberger, and J. Fong,
J. Cosmol. Astropart. Phys. 08 (2011)014;
E. McDonough and R. H. Brandenberger, 
J. Cosmol. Astropart.Phys. 02 (2013) 045.



\bibitem {neilsen}H. B. Nielsen and P. Olesen, 
Nucl. Phys. B 61 (1973) 45

\bibitem{abegrowth}D. Daverio, et al.
Physical Review D 93.8 (2016): 085014.

\bibitem{hartman}B. Hartmann, and P.J. Sirimachan,  High Energ. Phys. (2010) 2010: 110. 

\bibitem{wakes}J. Silk and A. Vilenkin, 
Phys. Rev. Lett.53, 1700 (1984); M. J. Rees,
Mon.Not. R. Astron. Soc.222, 27 (1986); T. Vachaspati, 
Phys.Rev. Lett.57, 1655 (1986); A. Stebbins, S. Veeraraghavan,
R. H. Brandenberger, J. Silk, and N. Turok, 
Astrophys. J.322, 1 (1987).

\bibitem{vachaspati} T,Vachaspati, 
Phys. Rev. D 45, 3487 (1992)


\bibitem{brihaye}Y. Brihaye and M. Lubo,
Phys. Rev. D 62 (2000) 085004

\bibitem{electroweak}A. Achucarro, and T. Vachaspati. 
Physics Reports 327.6 (2000): 347-426;
Holman, Richard, et al. 
Physical Review D 46.12 (1992): 5352.

\bibitem{ulbricht} S. Ulbricht and R. Meinel, Class. Quantum Grav. 32, 147001 (2015)

\bibitem {layek} B. Layek, S. Sanyal and A. M. Srivastava,  Phys. Rev.{\bf D 67}, 083508 (2003).

\bibitem{duple}F. Duplessis, and R. Brandenberger.
Journal of Cosmology and Astroparticle Physics 2013.04 (2013): 045.

\bibitem{cunha}da Cunha, Disrael Camargo Neves, R. H. Brandenberger, and O. F. Hernández. 
Physical Review D 93.12 (2016): 123501.


\end{thebibliography}
\end{document}